\documentclass[preprint]{revtex4}

\pdfoutput=1
\usepackage{graphicx,color}
\usepackage{amssymb}
\usepackage{amsmath}

\DeclareMathOperator{\Tr}{Tr}

\begin{document}
\title{Generalized activity equations for spiking neural network dynamics}

\author{Michael A. Buice$^1$}
\author{Carson C. Chow$^2$}
\affiliation{
$^1$Allen Institute for Brain Science\\
$^2$Laboratory of Biological  Modeling, NIDDK, NIH, Bethesda, MD}
\date{\today}

\begin{abstract}
Much progress has been made in uncovering the computational capabilities of spiking neural networks.  However, spiking neurons will always be more expensive to simulate compared to rate neurons because of the inherent disparity in time scales - the spike duration time is much shorter than the inter-spike time, which is much shorter than any learning time scale. In numerical analysis, this is a classic stiff problem. Spiking neurons are also much more difficult to study analytically.  One possible approach to making spiking networks more tractable is to augment mean field activity models with some information about spiking correlations.  For example, such a generalized activity model could carry information about spiking rates and correlations between spikes self-consistently.  Here, we will show how this can be accomplished by constructing a complete formal probabilistic description of the network and then expanding around a small parameter such as the inverse of the number of neurons in the network.  The mean field theory of the system gives a rate-like description.  The first order terms in the perturbation expansion keep track of covariances.
\end{abstract}

\maketitle

\section*{Introduction}

Even with the rapid increase in computing power due to Moore's law and proposals to simulate the entire human brain notwithstanding~\cite{markhram2012}, a realistic simulation of a functioning human brain performing nontrivial tasks is remote.  While it is plausible that a network the size of the human brain could be simulated in real time~\cite{izhikevich:2008ti,eliasmith:2012cz} there are no systematic ways to explore the parameter space.  Technology to experimentally determine all the parameters in a single brain simultaneously does not exist and any attempt to infer parameters by fitting to data would require exponentially more computing power than a single simulation. We also have no idea how much detail is required.  Is it sufficient to simulate a large number of single compartment neurons or do we need multiple-compartments? How much molecular detail is required? Do we even know all the important biochemical and biophysical mechanisms?  There are an exponential number of ways a simulation would not work and figuring out which remains computationally intractable.
Hence, an alternative means to provide appropriate prior distributions for parameter values and model detail is desirable.  Current theoretical explorations of the brain utilize either abstract mean field models or small numbers of more biophysical spiking models.  The regime of large but finite numbers of spiking neurons remains largely unexplored.  It is not fully known what role spike time correlations play in the brain.  It would thus be very useful if mean field models could be augmented with some spike correlation information.  

This paper outlines a scheme to derive generalized activity equations for the mean and correlation dynamics of a fully deterministic system of coupled spiking neurons. It synthesizes methods we have developed to solve two different types of problems. The first problem was how to compute finite system size effects in a network of coupled oscillators. We adapted the methods of the kinetic theory of gases and plasmas~\cite{Ichimaru:1973wn,Nicholson:1992} to solve this problem. The method exploits the exchange symmetry of the oscillators and characterizes the phases of all the oscillators in terms of a phase density function $\eta(\theta,t)$, where each oscillator is represented as a point mass in this density.  We then write down a formal flux conservation equation of this density,  called the Klimontovich equation, which completely characterizes the system. However, because the density is not differentiable, the Klimontovich equation only exists in the weak or distributional sense. Previously, e.g ~\cite{Desai:1978p8106,Strogatz:1991vw,Abbott:1993p938,Treves:1993p949} the equations were made usable by taking the ``mean field limit" of $N\rightarrow\infty$ and assuming that the density is differentiable in that limit, resulting in what is called the Vlasov equation. Instead of immediately taking the mean field limit, we regularize the density by averaging over initial conditions and parameters and then expand in the inverse system size $N^{-1}$ around the mean field limit. This results in a system of coupled moment equations known as the BBGKY moment hierarchy. In \cite{Hildebrand:2007el}, we solved the moment equations for the Kuramoto model perturbatively to compute the pair correlation function between oscillators. However, the procedure was somewhat {\it ad hoc} and complicated. We then subsequently showed in \cite{Buice:2007hla}, that the BBGKY moment hierarchy could be recast in terms of a density functional of the phase density. This density functional could be written down explicitly as an integral over all possible phase histories, i.e. a Feynman-Kac path integral. The advantage of using this density functional formalism is that the moments to arbitrary order in $1/N$ could be computed as a steepest-descent expansion of the path integral, which can be expressed in terms of Feynman diagrams. This made the calculation more systematic and mechanical. We later applied the same formalism to synaptically coupled spiking models~\cite{Buice:2013jw}.

Concurrently with this line of research, we also explored the question of how to generalize population activity equations, such as the Wilson-Cowan equations, to include the effects of correlations. The motivation for this question is that the Wilson-Cowan equations are mean field equations and do not capture the effects of spike-time correlations. For example, the gain in the Wilson-Cowan equations is fixed, (which is a valid approximation when the neurons fire asynchronously), but correlations in the firing times can change the gain~\cite{Salinas:2000}. Thus, it would be useful to develop a systematic procedure to augment population activity equations to include spike correlation effects. The approach we took was to posit plausible microscopic stochastic dynamics, dubbed the spike model,  that reduced to the Wilson-Cowan equations in the mean field limit and compute the self-consistent moment equations from that microscopic theory. Buice and Cowan~\cite{Buice:2009ts} showed that the solution of the master equation of the spike model could be expressed formally in terms of a path integral over all possible spiking histories. The random variable in the path integral is a spike count whereas in the path integral for the deterministic phase model we described above, the random variable is a phase density. To generate a system of moment equations for the microscopic stochastic system, we transformed the random spike count variable in the path integral into moment variables~\cite{Buice:2010p4742}. This is accomplished using the effective action approach of field theory, where the exponent of the cumulant generating functional, called the action, which is a function of the random variable is Legendre transformed into an effective action of the cumulants. The desired generalized Wilson-Cowan activity equations are then the equations of motion of the effective action. This is analogous to the transformation from Lagrangian variables of position and velocity to Hamiltonian variables of position and momentum. Here, we show how to apply the effective action approach to a deterministic system of synaptically coupled spiking neurons to derive a set of moment equations.

\section*{Approach}

Consider a network of single compartment conductance-based neurons
\begin{eqnarray*}
C\frac{dV_i}{dt}&=&-\sum_{r=1}^n g_r(x_i^r)(V_i-v_r)+\sum_{j=1}^N g_{ij} s_j(t)\\
\tau_i^r \frac{d x_i^r}{dt}&=&f(V_i,x_i)\\
\tau_j\frac{d s_j}{dt}&=&h(V_j,s_j)\\
\tau_g \frac{d g_{ij}}{dt} &=& \phi(g_{ij},V)
\end{eqnarray*}
The equations are remarkably stiff with time scales spanning orders of magnitude from milliseconds for ion channels, to seconds for adaptation, and from hours to years for changes in synaptic weights and connections.  Parameter values must be assigned for $10^{11}$ neurons with $10^4$ connections each.
Here, we present a formalism to derive a set of reduced activity equations directly from a network of deterministic spiking neurons that capture the spike rate and spike correlation dynamics.  The formalism first 
constructs a density functional for the firing dynamics of all the neurons in a network.  It then systematically marginalizes the unwanted degrees of freedom to isolate a set of self-consistent equations for the desired quantities.  For heuristic reasons, we derive an example set of generalized activity equations for the first and second cumulants of the firing dynamics of a simple spiking model but the method can be applied to any spiking model.

A convenient form to express spiking dynamics is with a phase oscillator.
Consider the quadratic integrate-and-fire neuron
\begin{equation}
\frac{dV_i}{dt} = I_i + V_i^2 + \alpha_i u(t)
\end{equation}
where $I$ is an external current and $u(t)$ are the synaptic currents with some weight $\alpha_i$.
The spike is said to occur when $V$ goes to infinity whereupon it is reset to minus infinity.  The quadratic nonlinearity ensures that this transit will occur in a finite amount of time.
The substitution $V=\tan(\theta/2)$ yields the theta model~\cite{Ermentrout:1986ve}:
\begin{equation}
\frac{d\theta_i}{dt}=1-\cos\theta_i + (1+\cos\theta_i)(I_i+\alpha_i u)
\end{equation}
which is the normal form of a Type I neuron near the bifurcation to firing~\cite{Ermentrout:1996wr}.  The phase neuron is an adequate approximation to spiking dynamics provided the inputs are not overly strong as to disturb the limit cycle.  The phase neuron also includes realistic dynamics such as not firing when the input is below threshold.  Coupled phase models arise naturally in weakly coupled neural networks~\cite{Ermentrout:1991wd,Golomb:2000p268,izhikevich:1997}.  They include the Kuramoto model~\cite{Kuramoto:1984}, which we have previously analyzed \cite{Hildebrand:2007el,Buice:2007hla}.  

Here, we consider the phase dynamics  of a set of $N$ coupled phase neurons obeying
\begin{eqnarray}
	\dot{\theta}_i &=& F(\theta,\gamma_i,u(t)) \label{eq:theta} \\
	\dot{u} (t)&=& -\beta u(t) +\beta \nu(t) \label{eq:drive}\\
	\nu(t)& =& \frac{1}{N}\sum_{j=1}^N\sum_l\delta(t-t^l_j) \label{eq:rate}
\end{eqnarray}
where each neuron has a phase $\theta_i$ that is indexed by $i$, $u$ is a global synaptic drive, $F(\theta,\gamma,u)$ is the phase and synaptic drive dependent frequency, $\gamma_i$ represents all the parameters for neuron $i$ drawn from a distribution with density $g(\gamma)$, $\nu$ is the  population firing rate of the network,$t^l_j$ is the $l$th firing time of neuron $j$ and a neuron fires when its phase crosses $\pi$.  
In the present paper, we consider all-to-all or global coupling through a synaptic drive  variable $u(t)$.  However, our basic approach is not restricted to global coupling.

 We can encapsulate the phase information of all the neurons into a neuron density function~\cite{Hildebrand:2007el,Buice:2007hla,Buice:2011tg,Buice:2013dt,Buice:2013jw}.
\begin{equation}
\eta(\theta,\gamma,t)=\frac{1}{N}\sum_{i=1}^N\delta(\theta-\theta_i(t))\delta(\gamma -\gamma_i)
\label{eq:empiricalMeasure}
\end{equation}
where $\delta(\cdot)$ is the Dirac delta functional, and $\theta_i(t)$ is a solution to system (\ref{eq:theta})-(\ref{eq:rate}).  The neuron density gives a count of the number of neurons with phase $\theta$ and synaptic strength $\gamma$ at time $t$.   
Using the fact that the Dirac delta functional in (\ref{eq:rate}) can be expressed as $\sum_l \delta(t-t_j^l)= \dot{\theta_j}\delta(\pi-\theta_j)$, the population firing rate can be rewritten as
\begin{equation}
	\nu(t)= \int d\gamma\,F(\pi,\gamma,u(t))\eta(\pi,\gamma,t)
	\label{eq:newrate}
\end{equation}

The neuron density formally obeys the conservation equation 
\begin{equation}
\frac{\partial}{\partial t} \eta (\theta, \gamma, t) + \frac{\partial}{\partial \theta} \left [ F \eta (\theta, \gamma, t) \right ]  =0
\label{eq:klimontovich}
\end{equation}
with initial condition $\eta(\theta,\gamma, t_0)=\eta_0(\theta,\gamma)$ and $u(t_0)=u_0$.  Equation (\ref{eq:klimontovich})
is known as the Klimontovich equation~\cite{Ichimaru:1973wn,Liboff:2003wn}.
The Klimontovich equation, the equation for the synaptic drive (\ref{eq:drive}), and the firing rate expressed in terms of the neuron density (\ref{eq:newrate}), fully define the system.  The system is still fully deterministic but is now in a form where various sets of reduced descriptions can be derived.
%In the classic population activity formalism, one can consider time or ensemble averages to produce a rate or activity variable.
Here, we will produce an example of a set of reduced equations or generalized activity equations that capture some aspects of the spiking dynamics.  The path we take towards the end will require the introduction of some formal machinery that may obscure the intuition around the approximations. However, we feel that it is useful because it provides a systematic and controlled way of generating averaged quantities that can be easily generalized. 

For finite $N$, (\ref{eq:klimontovich})  is only valid in the weak or distributional sense since $\eta$ is not differentiable.  In the $N\rightarrow\infty$ limit, it has been argued that $\eta$ will approach a smooth density $\rho$ that evolves according to the Vlasov equation that has the same form as  (\ref{eq:klimontovich}) but with $\eta$ replaced by $\rho$~\cite{Ichimaru:1973wn,Nicholson:1992,Desai:1978p8106,Strogatz:1991vw,Hildebrand:2007el}.  This has been proved rigorously in the case where noise is added using the theory of coupled diffusions~\cite{McKeanJr:1966wp,Faugeras:2009kn,Touboul:2012cv,Baladron:2011tg}.  This $N\rightarrow\infty$ limit is called mean field theory.  In mean field theory, the original microscopic many body neuronal network is represented by a smooth macroscopic density function. In other words, the ensemble of networks prepared with different microscopic initial conditions is sharply peaked at the mean field solution.  For large but finite $N$, there will be deviations away from  mean field~\cite{Hildebrand:2007el,Buice:2007hla,Buice:2013jw,Buice:2013dt}.  These deviations can be characterized in terms of a distribution over an ensemble of coupled networks that are all prepared with different initial conditions and parameter values.  Here, we show how a perturbation theory in $N^{-1}$ can be developed to expand around the mean field solution.  This requires the construction of the probability density functional over the ensemble of spiking neural networks.  We adapt the tools of statistical field theory to perform such a construction.

\subsection*{Formalism}
The complete description of the system given by equations (\ref{eq:drive}), ({\ref{eq:newrate}), and ({\ref{eq:klimontovich}) can be written as
\begin{align}
	\dot{u} (t)+\beta u(t) -\beta  \int d\gamma\,F(\pi,\gamma,u(t))\eta(\pi,\gamma,t)=0\label{eq:rate2}\\
\frac{\partial}{\partial t} \eta (\theta, \gamma, t) + \frac{\partial}{\partial \theta} \left [ F(\theta,\gamma,u(t)) \eta (\theta, \gamma, t) \right ] \equiv {\cal L}\eta =0
\label{eq:klim2}
\end{align}

The probability density functional governing the system specified by the  synaptic drive and Klimontovich equations  (\ref{eq:rate2}) and (\ref{eq:klim2}) given initial conditions $(\eta_0,u_0)$ can be written  as
\begin{align}
P[\eta,u]=&\int {\cal D} u_0(t)\, {\cal D}\eta_0(\theta,\gamma)\, P[\eta,u |\eta_0,u_0] \,P_0[\eta_0,u_0,\gamma]
\label{probdens}
\end{align}
where $P[\eta,u |\eta_0,u_0] $ is the conditional probability density functional of the functions $(\eta, u)$, and $P_0[\eta_0,u_0]$ is the density functional over initial conditions of the system. The integral is a Feynman-Kac path integral over all allowed initial condition functions.
Formally we can write $P[\eta,u |\eta_0,u_0] $ as a point mass (Dirac delta) located at the solutions of (\ref{eq:rate2}) and (\ref{eq:klim2}) given the initial conditions:
\begin{align*}
\delta\left[{\cal L} \eta -\eta_0\delta(t-t_0) \right]
\delta\left[\dot{u}+\beta u -\beta  \int d\gamma\,F(\pi,\gamma,u(t))\eta(\pi,\gamma,t)-u_0\delta(t-t_0)\right]
\end{align*}
The probability density functional (\ref{probdens}) is then
\begin{align}
P[\eta,u]=&\int {\cal D} u_0(t)\, {\cal D}\eta_0(\theta,\gamma)\, \delta\left[{\cal L} \eta -\eta_0\delta(t-t_0) \right] \nonumber\\
&\times\delta\left[\dot{u}+\beta u -\beta  \int d\gamma\,F(\pi,\gamma,u(t))\eta(\pi,\gamma,t)-u_0\delta(t-t_0)\right]P_0[\eta_0,u_0,\gamma]
\label{probdense2}
\end{align}

Equation (\ref{probdense2}) can be made useful by noting that the Fourier representation of a Dirac delta is given by $\delta(x)\propto \int dk\, e^{ikx}$.  Using the infinite dimensional Fourier functional transform then gives
$$
P[\eta,u] =  \int {\cal D} \tilde\eta {\cal D}\tilde u\,e^{-N S[\eta, \tilde{\eta}, u, \tilde{u}]}.
$$
The exponent $S[\eta,u]$ in the probability density functional is called the {\em action} and has the form
\begin{equation}
	S = S_u+ S_\varphi +S_0
	\label{eq:iafaction}
\end{equation}	
where 
\begin{eqnarray}
	S_\varphi= \int d\theta d\gamma dt  \,\tilde{\varphi}(x)\left [\partial_{t} \varphi(x)  + \partial_{\theta}F(\theta,\gamma,u(t)) \varphi(x)  \right ]  
	\label{eq:actionphi}
\end{eqnarray}
represents the contribution of the transformed neuron density to the action, 
\begin{eqnarray}
	S_u =\frac{1}{N}\int dt \,\tilde{u}(t) \left ( \dot u(t) + \beta u(t)   - \beta\int d\gamma F(\pi,\gamma,u(t)) [ \tilde{\varphi}(\pi,\gamma,t)  + 1]\varphi(\pi,\gamma,t)   \right ) 
	\label{eq:singleNeuronAction}
\end{eqnarray}
represents the global synaptic drive, $S_0[\tilde{\varphi}_0(x_0), u_0(t_0)] $ represents the initial conditions, and $x=(\theta,\gamma,t)$. For the case where the neurons are considered to be independent in the initial state, we have
\begin{eqnarray}
	S_0[\tilde{\varphi}_0(x_0), u_0(t_0)] &= -\frac{1}{N}\tilde{u}(t_0) u_0  - \ln \left ( 1 + \int d\theta d\gamma \tilde{\varphi}_0(\theta, \gamma, t_0) \rho_0 (\theta, \gamma, t_0) \right )
\end{eqnarray}
where $u_0$ is the initial value of the coupling variable and $\rho_0(\theta, \gamma, t)$ is the distribution from which the initial configuration is drawn for each neuron.
 The action includes two imaginary auxiliary response fields (indicated with a tilde), which are the infinite dimensional Fourier transform variables.  The factor of $1/N$ appears to ensure correct scaling between the $u$ and $\varphi$ variables since $u$ applies to a single neuron while $\varphi$ applies to the entire population. The full derivation is given in~\cite{Buice:2013jw} and a review of path integral methods applied to differential equations is given in \cite{Chow:2010}. In the course of the derivation we have made a Doi-Peliti-Jannsen transformation \cite{Janssen:2005wh,Buice:2013jw}, given by
\begin{eqnarray*}
	\varphi(x)&=& \eta(x) e^{-\tilde{\eta}(x)} \nonumber \\
	\tilde{\varphi}(x) &=&  e^{\tilde{\eta}(x)} - 1
\end{eqnarray*}
In deriving the action, we have explicitly chosen the Ito convention so that the auxiliary variables only depend on variables in the past. The action~(\ref{eq:iafaction}) contains all the information about the statistics of the network.  

The moments for this distribution can be obtained by taking functional derivatives of a moment generating functional.  Generally, the moment generating function for a random variable is given by the expectation value of the exponential of that variable with a single parameter. Because our goal is to transform to new variables for the first and second cumulants, we form a ``two-field" moment generating functional, which includes a second parameter for pairs of random variables, 
\begin{align}
\exp(N\, &W[J,K])=\nonumber\\
&\int {\cal D}\xi\, \exp\left[-NS[\xi] +N\int dx \, J^i(x) \xi_i(x) + \frac{N}{2} \int dx dx' \xi_i (x)K^{ij}(x,x') \xi_j(x') \right]
\label{eq:cgf}
\end{align}
where $J$ and $K$ are moment generating fields, $\xi_1(x)=u(t)$, $\xi_2(x)=\tilde u(t)$, $\xi_3(x)=\varphi(x)$, $\xi_4(x)=\tilde\varphi(x)$, and $x=(\theta,\gamma,t)$.
Einstein summation convention is observed beween upper and lower indices. Unindexed variables represent vectors. The integration measure $dx$ is assumed to be $dt$ when involving indices 1 and 2.  
%Indices can be raised or lowered by swapping a variable with its response variable, i.e. ($1\leftrightarrow 2$ and $3 \leftrightarrow 4$).  Hence $\xi^1=\tilde u$, $\xi^2=u$, $\xi^3=\tilde\varphi$, $\xi^4=\varphi$.
Covariances between an odd and even index corresponds to a covariance between a field and an auxiliary field.  Based on the structure of the action $S$ and (\ref{eq:cgf}) we see that this represents a linear propagator and by causality and the choice of the Ito convention is only nonzero if the time of the auxiliary field is evaluated at an earlier time than the field.  Covariances between two even indices correspond to that between two auxiliary fields and are always zero because of the Ito convention.

The mean and covariances of $\xi$ can be obtained by taking derivatives of the action $W[J,K]$ in (\ref{eq:cgf}), with respect to $J$ and $K$ and setting $J$ and $K$ to zero:
\begin{align*}
\frac{\delta W}{\delta J^i}&=\left.\langle \xi_i\rangle\right|_{J,K=0}\\
\frac{\delta W}{\delta K^{ij} }&=\left.\frac{1}{2}\langle \xi_i\xi_j\rangle\right|_{J,K=0}
\end{align*}
Expressions for these moments can be computed by expanding the path integral in (\ref{eq:cgf}) perturbatively around some mean field solution.  However, this can be unwieldy if closed form expressions for the mean field equations do not exist.
Alternatively, the moments at any order can be expressed as self-consistent dynamical equations that can be analyzed or simulated numerically.  Such equations form  a set of generalized activity equations for the means $a_i=\langle \xi_i\rangle$, and covariances
$C_{ij}=N[\langle \xi_i\xi_j\rangle -a_ia_j]$.  

We derive the generalized activity equations  by Legendre
 transforming the action $W$, which is a function of $J$ and $K$, to an effective action $\Gamma$ that is a function of $a$ and $C$. Just as a Fourier transform expresses a function in terms of its frequencies, a Legendre transform expresses a convex function in terms of its derivatives.  This is appropriate for our case because the moments are derivatives of the action.
The Legendre transform of $W[J,K]$ is
\begin{equation}
\Gamma[a,C]=-W[J,K]+\int dx J^i a_i + \frac{1}{2}\int dx dx' \left[a_i a_j+\frac{1}{N}C_{ij}\right]K^{ij}
\label{eq:lt}
\end{equation}
which must obey the constraints
\begin{align*}
\frac{\delta W}{\delta J^i}&=a_i\\
\frac{\delta W}{\delta K^{ij} }&=\frac{1}{2}\left[a_ia_j+ \frac{1}{N} C_{ij}\right]
\end{align*}
and
\begin{align}
\frac{\delta \Gamma}{\delta a_i}&\equiv \Gamma^{i,00}=J^i+\frac{1}{2}a_j\left[K^{ij}+K^{ji}\right]\nonumber\\
\frac{\delta \Gamma}{\delta C_{ij} }&\equiv\Gamma^{0,ij}=\frac{1}{2N}K^{ij}
\label{constraints}
\end{align}
The generalized activity equations are given by the equations of motion of the effective action, in direct analogy to the Euler-Lagrange equations of classical mechanics, and are obtained by setting $J^i=0$ and $K^{ij}=0$ in (\ref{constraints}).

In essence, what the effective action does is to take a probabilistic (statistical mechanical) system in the variables $\xi$ with action $S$ and transform them to a deterministic (classical mechanical) system with an action $\Gamma$.   Our approach here follows that used in  \cite{Buice:2010p4742} to construct generalized activity equations for the Wilson Cowan model.  However, there are major differences between that system and this one.  In \cite{Buice:2010p4742}, the microscopic equations were for the spike counts of an inherently probabilistic model so the effective action and ensuing generalized activity equations could be constructed directly from the Markovian spike count dynamics.
Here, we start from deterministically firing individual neurons and get to a probabilistic description through the Klimontovich equation.  It would be straightforward to include stochastic effects into the spiking dynamics.

Using (\ref{eq:lt}) in (\ref{eq:cgf}) gives
\begin{align}
\exp(-N\,\Gamma[a,C])=\int {\cal D}\psi\, \exp\left[-N S[\xi]+N\int dx \, J^i (\xi_i-a_i)\right.\nonumber\\
+ \left. \frac{N}{2}\int dx dx' \left[ \xi_i\xi_j - a_i a_j -\frac{1}{N}C_{ij}\right] K^{ij}\right]
\end{align}
where $J$ and $K$ are constrainted by (\ref{constraints}).
We cannot compute the effective action explicitly but we can compute it perturbatively in $N^{-1}$.  We first perform a shift $\xi_i = a_i+\psi_i$, expand the action as $S[a+\psi] = S[a]+\int dx(L^i[a]\psi_i + (1/2)\int dx' L^{ij}[a] \psi_i \psi_j) + \cdots$ and substitute for $J$ and $K$ with the constraints (\ref{constraints}) to obtain 
\begin{align}
\exp(-N\, \Gamma[a,C])=&\exp( -NS[a] -N\Tr \Gamma^{0,ij} C_{ij} )\int {\cal D}\psi\, \exp\left[-N\int dx \bigg (L^i[a]\psi_i \right.\nonumber\\
&\left.+\frac{1}{2}\int dx' L^{ij}[a] \psi_i \psi_j \bigg ) 
 + N\int dx \, \Gamma^{i,00} \psi_i + N^2\int dx dx' \psi_i\psi_j  \Gamma^{0,ij}\right]
\label{ea1}
\end{align}
where 
\begin{equation}
\Tr A^{ij}B_{ij} = \int dx dx' A^{ij}(x,x')B_{ij}(x,x')
\end{equation}

Our goal is to construct an expansion for $\Gamma$  by collecting terms in successive orders of $N^{-1}$ in the path integral of (\ref{ea1}).  Expanding  $\Gamma$ as $\Gamma[a,C] = \Gamma_0+N^{-1}\Gamma_1+N^{-2}\Gamma_2$ and equating coefficients of $N$ in (\ref{ea1}) immediately leads to the conclusion  that $\Gamma_0=S[a]$,  which gives
\begin{align}
\exp(-N\, \Gamma[a,C])=\exp\left( -NS[a] - \Tr \Gamma_1^{0,ij} C_{ij} \right)\int {\cal D}\psi\, \exp\left[-\frac{N}{2}\int dx L^{ij}[a] \psi_i \psi_j \right.\nonumber\\
+\left.N\int dx\, \Gamma_1^{0,ij}\psi_i\psi_j\right]\nonumber
\label{ea2}
\end{align}
where higher order terms in $N^{-1}$ are not included.  
To lowest nonzero order  $\Gamma^{0,ij}= N^{-1}\Gamma_1^{0,ij}$ since $\Gamma_0$ is only a function of $a$ and not $C$. If we set
\begin{equation}
 \Gamma_1^{0,ij}=(1/2) L^{ij} -(1/2)Q^{ij}, 
 \label{eq:ansatz}
 \end{equation}
 we obtain
\begin{align}
\exp(-N\, \Gamma[a,C])=\exp\left( -NS[a] - \frac{1}{2}\Tr L^{ij} C_{ij}+\frac{1}{2} \Tr Q^{ij} C_{ij} \right)\nonumber\\
\times \int {\cal D}\psi\, \exp\left[-\frac{N}{2}\int dx\, Q^{ij}[a] \psi_i \psi_j \right]
\label{ea3}
\end{align}
to order $N^{-1}$.  $Q^{ij}$ is an unknown function of $a$ and $C$, which we will deduce using self-consistency. 
The path integral in (\ref{ea3}), which is an infinite dimensional Gaussian that can be explicitly integrated, is proportional to $1/\sqrt{\det Q^{ij}}=\exp(-(1/2)\ln\det Q^{ij})=\exp(-(1/2)\Tr\ln Q^{ij})$, using properties of matrices.  Hence, (\ref{ea3}) becomes
\begin{align}
\exp(-N\, \Gamma[a,C])=\exp\left( -NS[a] - \frac{1}{2}\Tr L^{ij} C_{ij}-\frac{1}{2} \Tr Q^{ij} C_{ij} +\frac{1}{2}\Tr\ln Q^{ij} \right)\nonumber
\end{align}
and
\begin{align}
\Gamma[a,C]=S[a]  +\frac{1}{2N}\Tr L^{ij} C_{ij} +\frac{1}{2N}\Tr\ln Q^{ij} -\frac{1}{2N} \Tr Q^{ij} C_{ij}\nonumber
\end{align}
Taking the derivative of $\Gamma$ with respect to $C_{ij}$ yields
\begin{align}
\Gamma^{0,ij} = \frac{1}{2N}\left(L^{ij}+(Q^{-1})^{kl}\frac{\partial}{\partial C_{ij}} Q^{lk}-\frac{\partial}{\partial C_{ij}} (Q^{kl}C_{lk})\right)\nonumber
\end{align}
Self consistency with (\ref{eq:ansatz}) then requires that $Q^{ij}=(C^{-1})^{ij}$ which
leads to the effective action
\begin{equation}
\Gamma[a,C]=S[a]+\frac{1}{2N}\Tr\ln (C^{-1})^{ij}+\frac{1}{2N}\Tr L^{ij} C_{ij} 
\end{equation}
where 
$$\int dx'\, (C^{-1})^{ik}(x,x') C_{kj}(x',x_0) = \delta_{ij}\delta(x-x_0)$$
and we have dropped the irrelevant constant terms.

The equations of motion to order $N^{-1}$ are obtained from (\ref{constraints}) with $J^i$ and $K^{ij}$ set to zero:
\begin{align}
&\frac{\delta S[a]}{\delta a_i} + \frac{1}{2N}\frac{\delta}{\delta a_i} \Tr L^{ij} C_{ij}=0\label{mean}\\
&\frac{1}{2N}[-(C^{-1})^{ij}+L^{ij}]=0\label{corr1}
\end{align}
and (\ref{corr1}) can be rewritten as
\begin{align}
\int dx' L^{ik}(x,x')C_{kj}(x',x_0) = \delta_{ij}\delta(x-x_0)
\label{correlator}
\end{align}
Hence, given any network of spiking neurons, we can write down the a set of generalized activity equations for the mean and covariance functions by 1) constructing a neuron density function, 2) writing down the conservation law (Klimontovich equation), 3) constructing the action and 4) using formulas (\ref{mean}) and (\ref{correlator}).  We could have constructed these equations directly by multiplying the Klimontovich and synaptic drive equations by various factors of $u$ and $\eta$ and recombining.  However, as we saw in \cite{Buice:2010p4742} this is not a straightforward calculation.  The effective action approach makes this much more systematic and mechanical.

%Now we return to the solutions to the continuity equation which we separate into asynchronous and synchronous pieces:
%\begin{align}
%	\eta &\approx \rho + \delta \rho
%\end{align}
%and we insert this expansion into the continuity equation.  According with our intuition regarding the connection between ensemble averages and single network fluctuations, we must expand the continuity equation to at least second order in $\delta \rho$, as we expect the mean to be zero.

\subsection*{Phase Model Example}

We now present a simple example to demonstrate the concepts and approximations involved in our expansion.  Our goal is not to 
analyze the system {\it per se} but only to demonstrate the application of our method in a heuristic setting.
We begin with a simple nonleaky integrate-and-fire neuron model, which responds to a global coupling variable.  This is a special case of the dynamics given above, with $F$ given by
\begin{align}
	F[\theta, \gamma, u] = I(t) + \gamma u
\end{align}
The action from (\ref{eq:actionphi}) and (\ref{eq:singleNeuronAction}) is
\begin{align}
	S[a]=& \int d\theta d\gamma dt \,a_4(x)\left [\partial_{t} a_3(x)  + \partial_{\theta}(I+\gamma a_1(t)) a_3(x)  \right ] \nonumber \\
	&+\frac{1}{N}\int dt \,a_2(t) \left ( \dot a_1(t) + \beta a_1(t)   - \beta\int d\gamma\, (I+\gamma a_1(t)) [a_4(\pi,\gamma,t)  + 1]a_3(\pi,\gamma,t)   \right ) 
	\label{eq:actiona}
\end{align}
and we ignore initial conditions for now.

In order to construct the generalized activity equations we need to compute the first and second derivatives of the action $L^i$ and $L^{ij}$.  Taking the first derivative of (\ref{eq:actiona}) gives
\begin{align}
L^1[a](x,x')=\frac{\delta S[a(x)]}{\delta a_1(t')}&= \int d\theta d\gamma\,dt \gamma a_4(x) \partial_{\theta} a_3(x) \delta (t - t') \nonumber\\
&+\frac{1}{N}\left [\int dt\, a_2(t)\frac{d}{dt}\delta(t-t') +\beta a_2(t') - a_2(t')\beta\int d\gamma\, \gamma[a_4(\pi,\gamma,t')+1]a_3(\pi,\gamma,t')\right]\nonumber\\
L^2[a](x,x')=\frac{\delta S[a(x)]}{\delta a_2(t')} &= \frac{1}{N}\left[\frac{d a_1}{dt'} +\beta a_1(t') - \beta \int d\gamma(I+\gamma a_1(t'))[a_4(\pi,\gamma,t')+1]a_3(\pi,\gamma,t')\right]\nonumber\\
L^3[a](x,x')=\frac{\delta S[a(x)]}{\delta a_3(x')} &= \int dt\, a_4(\theta',\gamma',t)\partial_{t} \delta(t-t')+\int d\theta a_4(\theta,\gamma',t')\partial_\theta(I+\gamma' a_1(t')) \delta(\theta-\theta')\nonumber\\
& - \frac{\beta}{N} a_2(t') (I+\gamma' a_1(t'))(a_4(\pi,\gamma',t')+1)\delta(\pi-\theta')\nonumber\\
L^4[a](x,x')=\frac{\delta S[a(x)]}{\delta a_4(x')}&=\partial_{t'} a_3(x') +\partial_{\theta'}(I+\gamma' a_1(t')) a_3(x') -\frac{\beta}{N} a_2(t') (I+\gamma' a_1(t')) a_3(\pi,\gamma',t')\delta(\pi-\theta')\nonumber\\
\label{firstderivs}
\end{align}
The mean field equations are obtained by solving $L^{i}=0$ using (\ref{firstderivs}).  We immediately see that $a_2=a_4=0$ are solutions, which leaves us with
\begin{align}
\dot{a_1} +\beta a_1 - \beta \int d\gamma(I+\gamma a_1)a_3(\pi,\gamma,t)=0\label{drive}\\
\partial_t a_3 +(I+\gamma a_1)\partial_\theta a_3 =0\label{vlasov}
\end{align}
The mean field equations should be compared to those of the spike response model~\cite{Gerstner:1995p643,Gerstner:2000p665}.
We can also solve (\ref{vlasov}) directly to obtain
\begin{equation*}
	a_3(x, t) = \rho_0 \left (\theta - \int_{t_0}^t dt' \left [   I_\Omega(t') + \gamma  a_1(t')\right ] , \gamma, \Omega \right ) 
\end{equation*}
where $\rho_0$ is the initial distribution. If the neurons are initially distributed uniformly in phase, then $\rho_0 = g(\gamma)/2\pi$ and the mean field equations reduce to
\begin{equation}
\dot{a_1}(t) + \beta a_1(t) - \frac{\beta}{2\pi}   \left (  I + \bar\gamma  a_1(t) \right ) = 0
\end{equation}
which has the form of the Wilson-Cowan equation, with $(\beta/2\pi) \left (  I + \bar\gamma  a_1 \right )$ acting as a gain function.  Hence, the Wilson-Cowan equation is a full description of the infinitely large system limit of a network of globally coupled simple phase oscillators in the asynchronous state.  For all other initial conditions,  the one-neuron conservation equation (called the Vlasov equation in kinetic theory) must be included in mean field theory.

To go beyond mean field theory we need to compute  $L^{ij}(x,x',x'')=\delta L^{i}(x,x')/\delta a_j(x'')$:
\begin{align*}
L^{11}[a]&=  0\\
L^{12}[a]&= \frac{1}{N}\left [-\frac{d}{dt''} +\beta-\beta\int d\gamma\, \gamma[a_4(\pi,\gamma,t'')+1]a_3(\pi,\gamma,t'')]\right]\delta(t''-t')\\
L^{13}[a]&= \left[\gamma'' \int d\theta\, a_4(x) \delta (\gamma - \gamma'') \partial_\theta \delta(\theta-\theta'') -\frac{\beta}{N} \gamma'' a_2(t')[a_4(\pi,\gamma'',t'')+1]\delta(\pi-\theta'')\right]\delta(t'-t'')\\
L^{14}[a]&=  \left[\gamma'' \partial_{\theta''} a_3(x'') -\frac{\beta}{N}\gamma'' a_2(t'') a_3(\pi,\gamma'',t'')\delta(\pi-\theta'')\right]\delta(t'-t'')
\end{align*}
\begin{align*}
L^{21}[a]&= \frac{1}{N}\left[\frac{d}{dt'} +\beta- \beta\int d\gamma\, \gamma[a_4(\pi,\gamma,t')+1]a_3(\pi,\gamma,t')\right]\delta(t'-t'') \\
L^{22}[a]&=  0\\
L^{23}[a]&=  -\frac{\beta}{N}  (I+\gamma'' a_1(t'))[a_4(\pi,\gamma'',t'))+1]\delta(\pi-\theta')\delta(t'-t'')\\
L^{24}[a]&=  -\frac{\beta}{N} (I+\gamma'' a_1(t'))a_3(\pi,\gamma'',t')]\delta(\pi-\theta'')\delta(t'-t'')
\end{align*}
\begin{align*}
L^{31}[a]& = \left[\int d\theta\, a_4(\theta,\gamma',t') \gamma'\partial_\theta \delta(\theta-\theta')-\frac{\beta}{N}a_2(t')\gamma'[a_4(\pi,\gamma',t')+1]\delta(\pi-\theta')\right]\delta(t'-t'')\\
L^{32}[a]& =  -\frac{\beta}{N}(I+\gamma'a_1(t'))(a_4(\pi,\gamma',t')+1)\delta(\pi-\theta')\delta(t'-t'')\\
L^{33}[a]& =0\\
L^{34}[a]& = \left[\delta(\theta'-\theta'')\partial_{t''}-\partial_{\theta''}(I+\gamma' a_1(t'))\delta(\theta''-\theta') \right.\\
&\left.- \frac{\beta}{N} a_2(t') (I+\gamma' a_1(t'))\delta(\pi-\theta')\delta(\pi-\theta'')\right]\delta(\gamma'-\gamma'')\delta(t''-t')
\end{align*}
\begin{align*}
L^{41}[a]&= \left[\partial_{\theta'}\gamma'a_3(x')-\frac{\beta}{N}a_2(t')\gamma'a_3(\pi,\gamma',t')\delta(\pi-\theta')\right]\delta(t'-t'')\\
L^{42}[a]&= -\frac{\beta}{N}(I+\gamma'a_1(t'))a_3(\pi,\gamma',t')\delta(\pi-\theta')\delta(t'-t'')\\
L^{43}[a]&=\partial_{t'}\delta(x'-x'')  + \partial_{\theta'}(I+\gamma' a_1(t'))\delta(x'-x'')  \\
&- \frac{\beta}{N} a_2(t')(I+\gamma a_1(t'))\delta(\pi-\theta')\delta(\pi-\theta'')\delta(\gamma'-\gamma'')\delta(t'-t'')\\
L^{44}[a]&=0
\end{align*}

The activity equations for the means to order $N^{-1}$ are given by (\ref{mean}).  The only nonzero contributions are given by
$L^{13}$ and $L^{31}$ resulting in 
\begin{align*}
&L^2+\frac{1}{2N}\frac{\delta}{\delta a_2} \int dx dx' (L^{13}C_{13}+L^{31}C_{31})=0\\
&L^4+\frac{1}{2N}\frac{\delta}{\delta a_4} \int dx dx' (L^{13}C_{13}+L^{31}C_{31})=0
\end{align*}
since $a_2=a_4=0$ and correlations involving response variables (even indices) will be zero for equal times. The full activity equations for the means are thus
\begin{align}
&\dot{a_1} +\beta a_1 - \beta \int d\gamma(I+\gamma a_1)a_3(\pi,\gamma,t) - \frac{\beta}{N}\int d\gamma\, \gamma C(\pi,\gamma,t)=0\\
&\partial_t a_3 +(I+\gamma a_1)\partial_\theta a_3 + \frac{1}{N}\gamma \partial_\theta C(\theta,\gamma,t)=0
\end{align}
where $C(\theta,\gamma,t)=C_{13}(t;\theta,\gamma,t)=C_{31}(\theta,\gamma,t;t)$.

We can now use the $L^{ij}$ in (\ref{correlator}) to obtain activity equations for $C_{ij}$.  There will be sixteen coupled equations in total but the applicable nonzero ones are
%\begin{align}
%1,1&L^{12}C_{21}+L^{14}C_{41}=\delta\\
%1,2&L^{12}C_{22}+L^{14}C_{42}=0\\
%1,3&L^{12}C_{23}+L^{14}C_{43}=0\\
%1,4&L^{12}C_{24}+L^{14}C_{44}=0\\
%2,1&L^{21}C_{11}+L^{23}C_{31}+L^{24}C_{41}=0\\
%2,2&L^{21}C_{12}+L^{23}C_{32}+L^{24}C_{42}=\delta\\
%2,3&L^{21}C_{13}+L^{23}C_{33}+L^{24}C_{43}=0\\
%2,4&L^{21}C_{14}+L^{23}C_{34}+L^{24}C_{44}=0\\
%3,1&L^{32}C_{21}+L^{34}C_{41}=0 \\
%3,2&L^{32}C_{22}+L^{34}C_{42}=0\\
%3,3&L^{32}C_{23}+L^{34}C_{43}=\delta\\
%3,4&L^{32}C_{21}+L^{34}C_{41}=0\\
%4,1&L^{41}C_{11}+L^{42}C_{21}+L^{43}C_{31}=0\\
%4,2&L^{41}C_{12}+L^{42}C_{22}+L^{43}C_{32}=0\\
%4,3&L^{41}C_{13}+L^{42}C_{23}+L^{43}C_{33}=0\\
%4,4&L^{41}C_{14}+L^{42}C_{24}+L^{43}C_{34}=\delta
%\end{align}
\begin{align}
%2,1&L^{21}C_{11}+L^{23}C_{31}+L^{24}C_{41}=0\\
%2,3&L^{21}C_{13}+L^{23}C_{33}+L^{24}C_{43}=0\\
%4,1&L^{41}C_{11}+L^{42}C_{21}+L^{43}C_{31}=0\\
%4,3&L^{41}C_{13}+L^{42}C_{23}+L^{43}C_{33}=0\\
\left [\frac{d}{dt} +\beta- \beta\int d\gamma\, \gamma a_3(\pi,\gamma,t)]\right]&C_{11}(t;t_0)-\beta
\int d\gamma \, (I+\gamma a_1)C_{31}(\pi,\gamma,t;t_0)\nonumber\\
&-\beta\int d\gamma\, (I+\gamma a_1(t))a_3(\pi,\gamma,t)C_{41}(\pi,\gamma,t;t_0)=0\label{first}\\
\left[\frac{d}{dt} +\beta- \beta\int d\gamma\, \gamma a_3(\pi,\gamma,t)\right]&C_{13}(t;x_0)-\beta  \int d\gamma\,(I+\gamma a_1)C_{33}(\pi,\gamma,t;x_0)\nonumber\\
&-\beta\int d\gamma\, (I+\gamma a_1(t))a_3(\pi,\gamma,t)C_{43}(\pi,\gamma,t;x_0)=0\label{second}\\
\gamma \partial_\theta a_3(x)C_{11}(t;t_0) +[\partial_t  +&(I+\gamma a_1)\partial_\theta] C_{31}(x;t_0) \nonumber\\
&-\frac{\beta}{N} (I+\gamma a_1(t))a_3(\pi,\gamma,t)\delta(\pi-\theta)C_{21}(t,t_0)=0\label{third}\\
\gamma \partial_\theta a_3(x) C_{13}(t;x_0)  +[\partial_t  +&(I+\gamma a_1(t))\partial_\theta ] C_{33}(x,x_0) \nonumber    \\
&-  \frac{\beta}{N} (I+\gamma a_1(t))a_3(\pi,\gamma,t)\delta(\pi-\theta)C_{23}(t,x_0)=0\label{fourth}
\end{align}

Adding (\ref{second}) and (\ref{third}) and taking the limit $t_0\rightarrow t$ and setting $\theta_0=\theta$, $\gamma_0=\gamma$ gives
\begin{align*}
\partial_t &C(\theta,\gamma,t)+\left[\beta- \beta\int d\gamma'\, \gamma' a_3(\pi,\gamma',t)+(I+\gamma a_1)\partial_\theta\right]C(\theta,\gamma,t)-\beta  \int d\gamma'\,(I+\gamma' a_1)C_{33}(\pi,\gamma',t;x)\nonumber\\
&-2\beta (I+\gamma a_1(t))a_3(\pi, \gamma, t)\delta(\pi-\theta)+\gamma \partial_\theta a_3(x)C_{11}(t;t)  =0 %-\beta (I+\gamma a_1(t))a_3(\pi,\gamma,t)\delta(\pi-\theta)=0
\end{align*}
where we use the fact that $C_{21}(t,t') = N$ and $C_{43}(x;x') = \delta(\theta-\theta')\delta(\gamma-\gamma')$ in the limit of $t'$ approaching $t$ from below and equal to zero when approaching from above.  Adding (\ref{first}) and (\ref{fourth})  to themselves with $t$ and $t_0$ interchanged and taking the limit of $t_0$ approaching $t$ gives
\begin{align*}
&\left [\frac{d}{dt} +2\beta- 2\beta\int d\gamma\, \gamma a_3(\pi,\gamma,t)]\right]C_{11}(t;t)-2\beta
\int d\gamma \, (I+\gamma a_1)C(\pi,\gamma,t)=0\\
%&-\beta\int d\gamma\, (I+\gamma a_1(t))a_3(\pi,\gamma,t)=0\\
&\left[\partial_t  +(I+\gamma a_1(t))\partial_\theta  \right] C_{33}(x;x) +2\gamma[ \partial_\theta a_3(x)] C(x)=0 
%&- \beta (I+\gamma a_1(t))a_3(\pi,\gamma,t)\delta(\pi-\theta)=0
\end{align*}
because  $C_{41}(x;t) = 0$ and $C_{23}(t;x) = 0$.
Putting this all together, we get the generalized activity equations
\begin{align}
&\frac{{da_1}}{dt}+\beta a_1(t) - \beta \int d\gamma(I+\gamma a_1(t))a_3(\pi,\gamma,t) - \frac{\beta}{N}\int d\gamma\, \gamma C(\pi,\gamma,t)=0\label{a1}\\
&\partial_t a_3(\theta,\gamma,t) +(I+\gamma a_1)\partial_\theta a_3(\theta,\gamma,t) + \frac{1}{N} \gamma  \partial_\theta C(\theta,\gamma,t)=0
\label{2nd}\\
&\partial_t C(\theta,\gamma,t)+\left[\beta- \beta\int d\gamma'\, \gamma' a_3(\pi,\gamma',t)+(I+\gamma a_1)\partial_\theta\right]C(\theta,\gamma,t)\nonumber\\
&-\beta  \int d\gamma'\,(I+\gamma' a_1)C_{33}(\pi,\gamma',t;\theta,\gamma,t)-2\beta  (I+\gamma a_1(t))a_3(\theta,\gamma,t)\delta(\pi-\theta)\nonumber\\
&+\gamma \partial_\theta a_3(\theta,\gamma,t)C_{11}(t;t)= 0 \\%-\beta (I+\gamma a_1(t))a_3(\pi,\gamma,t)\delta(\pi-\theta)=0\\
&\left [\frac{d}{dt} +2\beta- 2\beta\int d\gamma\, \gamma a_3(\pi,\gamma,t)]\right]C_{11}(t;t)-2\beta
\int d\gamma \, (I+\gamma a_1)C(\pi,\gamma,t) =0\\
%&-\beta\int d\gamma\, (I+\gamma a_1(t))a_3(\pi,\gamma,t)=0\\
&\left[\partial_t  +(I+\gamma a_1(t))\partial_\theta  \right] C_{33}(\theta,\gamma,t;\theta,\gamma,t) +2\gamma \partial_\theta a_3(\theta,\gamma,t) C(\theta,\gamma,t)=0\label{C33}
% &- \beta (I+\gamma a_1(t))a_3(\pi,\gamma,t)\delta(\pi-\theta)=0
\end{align}
Initial conditions, which are specified in the action, are required for each of these equations.  The derivation of these equations using classical means require careful consideration for each particular model.  Our method provides a blanket mechanistic algorithm. We propose that these equations represent a new scheme for studying neural networks.

Equations (\ref{a1})-(\ref{C33}) are the complete self-consistent generalized activity equations for the mean and correlations to order $N^{-1}$.
It is a system of partial differential equations in $t$ and $\theta$.  These equations can be directly analyzed or numerically simulated.
Although the equations seem complicated, one must bear in mind that they represent the dynamics of the system averaged over initial conditions and unknown parameters.  Hence, the solution of this PDE system replaces multiple simulations of the original system. In previous work, we required over a million simulations of the original system to obtained adequate statistics~\cite{Buice:2013jw}. There is also a possibility that simplifying approximations can be applied to such systems. The system has complete phase memory because the original system was fully deterministic. However, the inclusion of stochastic effects will shorten the memory and possibly simplify the dynamics. It will pose no problem to include such stochastic effects. In fact, the formalism is actually more suited for stochastic systems~\cite{Buice:2010p4742}.

\section*{Discussion}
The main goal of this paper was to show how to systematically derive generalized activity equations for the ensemble averaged moments of a deterministically coupled network of spiking neurons. Our method utilizes a path integral formalism that makes the process algorithmic.  The resulting equations could be derived using more conventional perturbative methods although possibly with more calculational difficulty as we found before~\cite{Buice:2010p4742}.  For example, for the case of the stochastic spike model, Buice et al.~\cite{Buice:2010p4742} presumed that the Wilson-Cowan activity variable was the rate of a Poisson process and derived a system of generalized activity equations that corresponded to deviations around Poisson firing. 
Bressloff~\cite{Bressloff:2010jc}, on the other hand, assumed that the Wilson-Cowan activity variable was a mean density and 
used a system-size expansion to derive an alternative set of generalized activity equations for the spike model. The classical derivations of these two interpretations look quite different and the differences and similarities between them are not readily apparent.  However, the connections between the two types of expansions are very transparent using the path integral formalism.

Here, we derived equations for the rate and covariances (first and second cumulants) of a deterministic synaptically coupled spiking network as a system size expansion to first order.  However, our method is not restricted to these choices. What is particularly advantageous about the path integral formalism is that it is straightforward to generalize to include higher order cumulants, extend to higher orders in the inverse system size, or to expand in other small parameters such as the inverse of a slow time scale.  The action fully specifies the system and all questions regarding the system can be addressed with it.

To give a concrete illustration of the method, we derived the self-consistent generalized activity equations for the rates and covariances to order $N^{-1}$  for a simple phase model.  The resulting equations consist of ordinary and partial differential equations.  This is to be expected since the original system was fully deterministic and memory cannot be lost.  Even mean field theory requires the solution of an advective partial differential equation. The properties of these and similar equations remain to be explored computationally and analytically.
The system is possibly simpler near the asynchronous state, which is marginally stable in mean field theory like the Kuramoto model~\cite{Strogatz:1991vw} and like the Kuramoto model, we conjecture that the finite size effects will stabilize the asynchronous state~\cite{Hildebrand:2007el,Buice:2007hla}.  The addition of noise will also stabilize the asynchronous state.  Near asynchrony could be exploited to generate simplified versions of the asynchronous state.  
%We gave an example although the simplification was perhaps too drastic and resulted in a linear system. 

We considered a globally connected network, which allowed us to assume that networks for different parameter values and initial conditions converge towards a ``typical" system in the large $N$ limit.  However, this property may not hold for more realistic networks.  While the formalism describing the ensemble average will hold regardless of this assumption, the utility of the equations as descriptions of a particular network behavior may suffer.  For example, heterogeneity in the connectivity (as opposed to the global connectivity we consider here) may threaten this assumption.  This is the case with so called ``chaotic random networks" \cite{Sompolinsky:1988p1714} in which there is a spin-glass transition owing to the variance of the connectivity crossing a critical threshold.  This results in the loss of a ``typical" system in the large $N$ limit requiring an effective stochastic equation which incorporates the noise induced by the network heterogeneity.  Whether the expansion we present here is useful without further consideration depends upon whether the network heterogeneity induces this sort of effect.  This is an area for future work.  A simpler issue arises when there are a small discrete number of ``typical" systems (such as with bistable solutions to the continuity equation).  In this case, there are noise induced transitions between states.  While the formalism has a means of computing this transition \cite{Elgart:2004vi}, we do not consider this case here.  

An alternative means to incorporate heterogeneous connections is to consider a network of coupled systems.  In such a network, a set of generalized activity equations, such as those derived here or simplified versions, would be derived for each local system, together with equations governing the covariances between the local systems.  Correlation based learning dynamics could then be imposed on the connections between the local systems. Such a network could serve as a generalization of current rate based neural networks to include the effects of spike correlations with applications to both neuroscience and machine learning.

\section*{Acknowledgments}
This work was supported by the Intramural Research Program of the NIH, NIDDK.

\bibliography{pcbRefs}

\end{document}